\documentclass[letterpaper]{article}

\usepackage[T1]{fontenc}

\usepackage{geometry}
\usepackage{setspace}
\usepackage{lineno}

\usepackage[
  style = chem-acs,
  articletitle = true,
  doi = true,
  url = false,
  maxbibnames = 15
]{biblatex}
\usepackage{graphicx}
\usepackage{float}
\newfloat{scheme}{htbp}{los}
\floatname{scheme}{Scheme}
\newfloat{chart}{htbp}{loc}
\floatname{chart}{Chart}
\newfloat{graph}{htbp}{loh}
\floatname{graph}{Graph}

\usepackage{amsmath,amssymb}
\usepackage{booktabs}
\usepackage{siunitx}
\usepackage{xcolor}
\usepackage[hidelinks]{hyperref}

\usepackage{authblk}
\author[1]{Charles Dove*}
\author[1]{Laura Waller*}
\affil[1]{Department of Electrical Engineering and Computer Sciences, University of California, Berkeley, Berkeley, California 94720, United States}

\title{Inductively Scalable, Single-Step Neural Surrogates for Wave-Scattering Inverse Problems}

\date{*Email: charles\_dove@berkeley.edu; waller@berkeley.edu}

\begin{document}

\maketitle

\begin{abstract}
Neural network surrogates are an emerging alternative to traditional electromagnetic wave simulators like finite-difference time-domain (FDTD); their goal is to replace rigorous physical simulations with pre-trained neural networks that solve wave-scattering forward and inverse problems orders of magnitude faster. However, nonrecurrent, single-step surrogates have scaled only to a few tens of simulation variables. Here, we show that this barrier can be overcome by dynamically generating salient training examples during training, rather than randomly sampling the large space of possible examples. We introduce an algorithm that runs in parallel with surrogate training, using gradient ascent to search refractive-index and source configurations for cases where the surrogate disagrees with a full-wave ground-truth simulator. We also use source and ground-truth normalization with an evolving replay dataset to stabilize and accelerate learning. Using this approach, we train a fast, single-step surrogate for two-dimensional wave scattering with up to 41,772 controllable variables, including dense, freely configurable grids of refractive indices and complex-valued sources. The resulting neural surrogate is robustly accurate across diverse structured and unstructured examples and generalizes inductively to larger domains, reaching over 3 million controllable variables without retraining, a $73.8\times$ increase. We demonstrate the surrogate on large-scale forward simulations and inverse design of freeform beam splitters and gradient-index (GRIN) lenses up to 98 wavelengths wide, showing comparable or better performance than FDTD-based designs, with speedups from $1.29\times$ to $26.5\times$. These results demonstrate a practical path toward fast, robustly accurate, inductively scalable neural simulators for photonic inverse design and other wave-scattering inverse problems.
\end{abstract}

\section*{Keywords}
neural surrogate; wave scattering; inverse design; photonics; finite-difference time-domain; dynamic hard-example generation; gradient-index optics

\section*{Abbreviations}
FDTD, finite-difference time-domain; GRIN, gradient-index; NN, neural network; PML, perfectly matched layer; RMS, root mean square; TE, transverse electric

\section{Introduction}
Simulating the interaction between electromagnetic waves and structured materials is fundamental to a wide range of light-based science and engineering, including lens design, photonic device optimization~\cite{Fan2020}, and computational imaging~\cite{Eckert2019}. In many of these settings, the forward simulation must be differentiated and repeated thousands to millions of times to solve a given inverse problem. For example, in photonic inverse design, we iteratively optimize a refractive index map to find the one that best performs the desired function. At each iteration, we must run a forward model to simulate the input wave propagating through the device~\cite{Molesky2018,LalauKeraly2013,Hughes2019,Minkov2020}. High-fidelity full-wave solvers such as the finite-difference time-domain (FDTD) method are the gold standard, but remain computationally expensive at the resolutions and domain sizes required for realistic devices, and the associated memory and runtime costs are often infeasible for such iterative design methods~\cite{Yee1966,Taflove2005,Oskooi2010}. 

Neural surrogate models offer a compelling alternative: rather than solving Maxwell's equations from scratch for every new configuration, a neural network can be trained to approximate the mapping from a given scattering problem instance to the resulting electric field distribution~\cite{Jiang2020,Ma2020,Chen2022WaveYNet,Lim2022MaxwellNet,Augenstein2023,Pestourie2020ActiveLearning}, ideally at a speed much faster than that achieved by FDTD. Specifically, a neural surrogate will take as input a set of simulation variables (e.g. a 2D map of refractive index values) and a specification of the wave sources in the region. The simulation variables are usually defined on a grid or as a discrete set of points, or described with geometric parameters, such as the individual widths of a set of nanostrips~\cite{Liu2018}. The neural surrogate then predicts the electric field, magnetic field, and/or intensity distribution in the region, at either a full region-covering grid or at one or more points. This approximate forward mapping may be differentiated and used in place of a traditional forward simulator, such as FDTD, the finite difference frequency domain (FDFD) method, or Green's function based methods, when solving inverse problems.

There are two primary approaches to neural surrogates: single-step "direct" solvers, which are fast but small scale, and iterative solvers, which are larger-scale but slow. Iterative solvers repeatedly or recurrently apply a neural network many times to perform a single simulation (which may then be inserted into a larger iterative loop to solve an inverse problem). Substantial progress has been made in recent years on improving the spatial scaling and accuracy of iterative neural surrogates for wave scattering problems. One line of work uses domain-decomposition methods, which tile larger regions from small subdomains~\cite{Mao2024,Mao2026}. A separate line of work uses neural-network-based preconditioners to accelerate traditional iterative solution methods~\cite{Rizzuti2019Helmholtz,Azulay2022HelmholtzPreconditioner,Lerer2024HelmholtzPreconditioner}. These iterative methods, however, typically require tens or hundreds of sequential neural-network inferences \textit{within  a single simulation}, which must then be repeated thousands to millions of times to solve a given inverse problem. This many-step recurrence issue typically scales with the spatial size of the simulation; for instance, a domain-decomposition method, for a domain that is $w$ tiles wide, requires at least $w$ sequential inference rounds to propagate information across the domain, and high refractive index contrast simulations may require substantially more than this due to resonance effects. 

\begin{figure}[]
\centering\includegraphics[width=\textwidth]{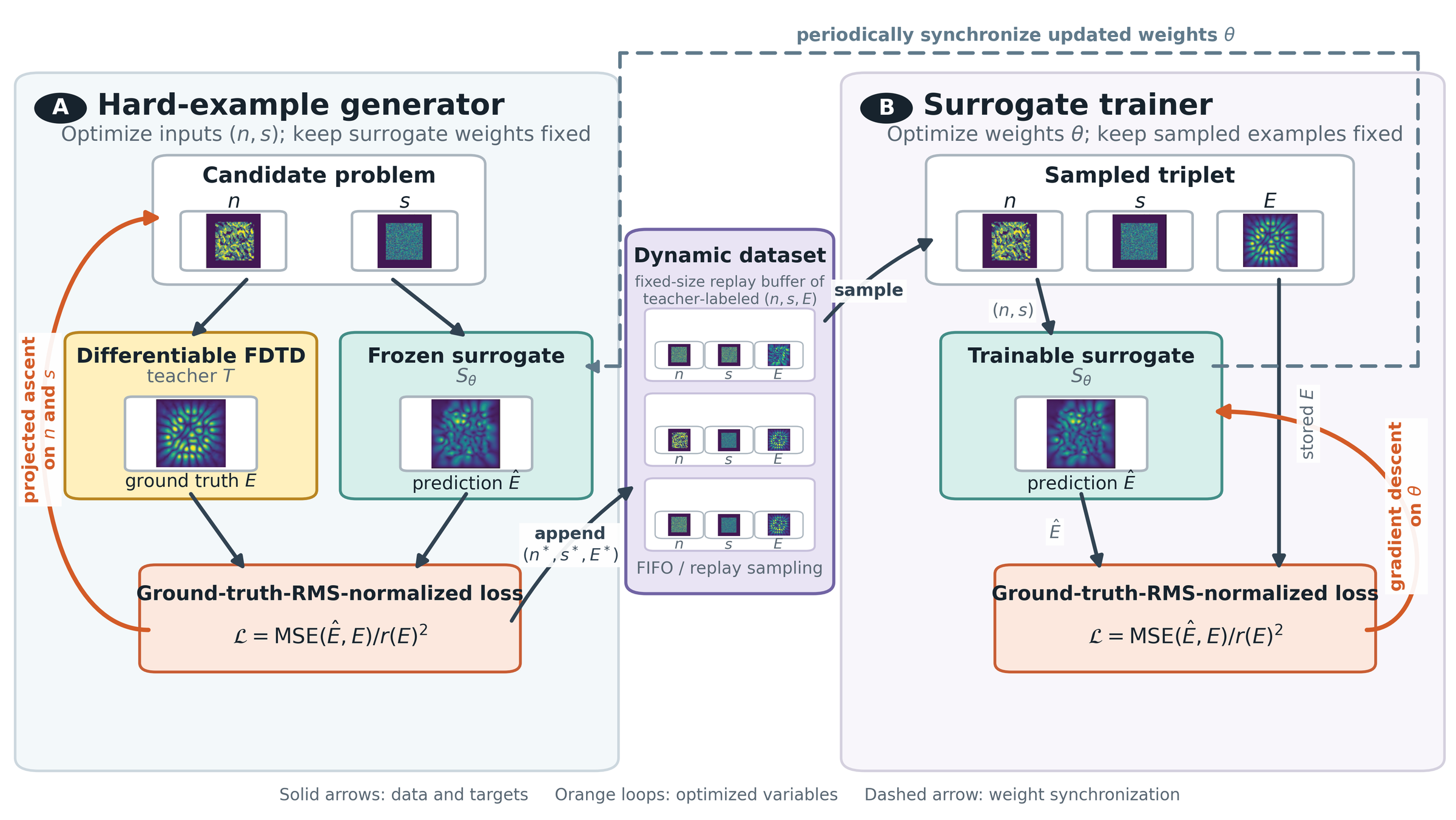}
\caption{\textbf{Workflow for dynamic training data generation and neural surrogate training}: (A) A dynamic dataset generation process searches for training data examples that maximize disagreement between the current neural surrogate and a ground-truth Finite-Difference Time-Domain (FDTD) output. These high-loss examples (consisting of refractive-index (n) and source (s) configurations) where the neural surrogate performs poorly are then added to the dynamic training dataset, along with their ground-truth FDTD output (E). The neural surrogate is periodically replaced by its updated version from the bottom process. (B) In parallel, the dynamic dataset elements are used to train the neural surrogate. As the surrogate improves, the dataset generation process discovers new failure modes, creating increasingly informative training examples.}
\label{diagram}
\end{figure}

\begin{figure}[]
\centering\includegraphics[width=15cm]{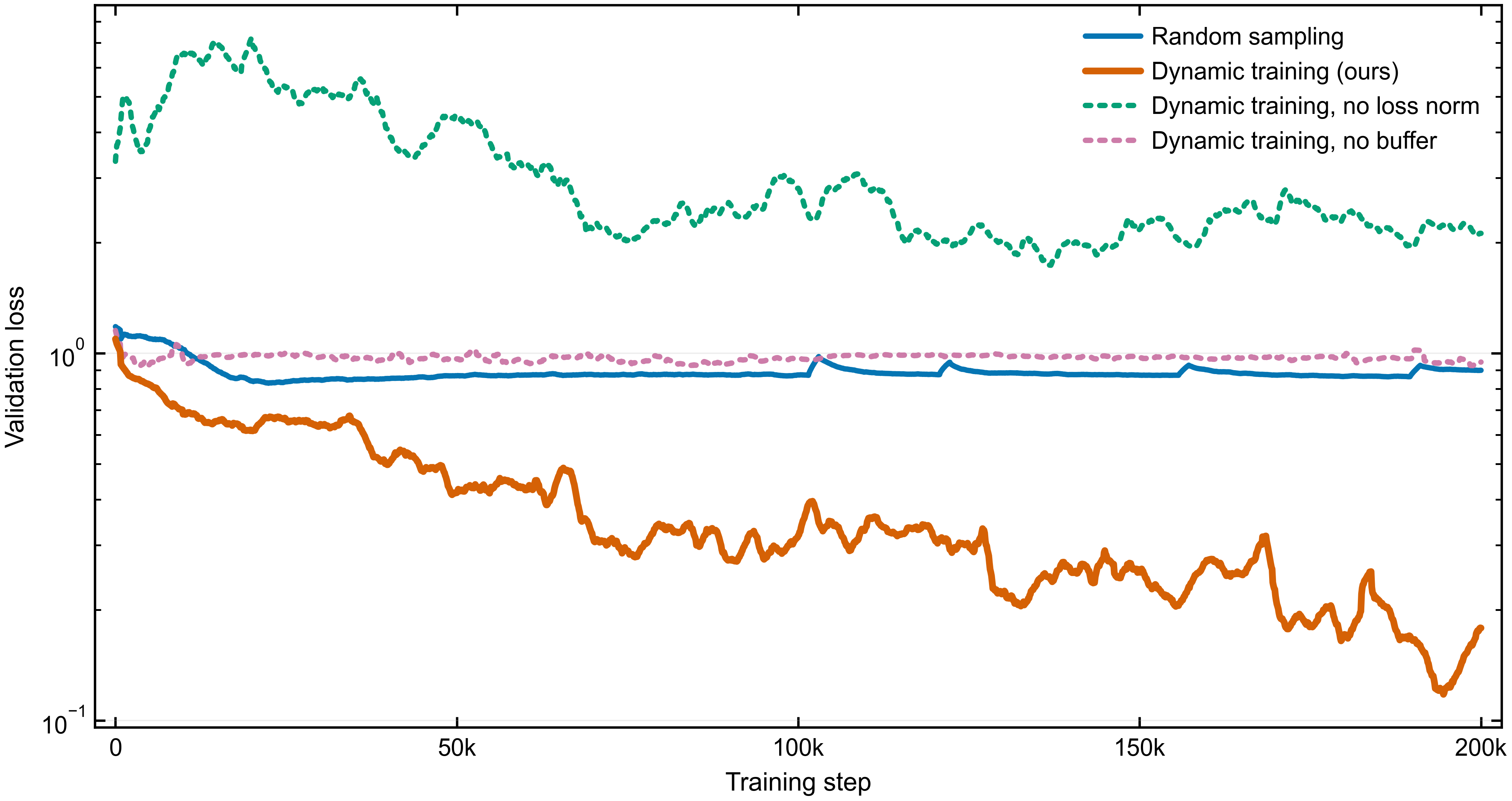}
\caption{Our dynamic training approach (orange) performs significantly better than random sampling of the training examples (blue), with respect to validation loss. We also demonstrate the importance of each component of our method by showing results for our proposed dynamic sampling method without the loss normalization (green) and without using a buffer of training examples (pink), i.e. not using the dynamic dataset and instead training immediately on generated examples.}
\label{fig:dynamic_vs_random_loss}
\end{figure}

Single-step solvers directly predict the output of the simulation, typically the electric field distribution~\cite{Jiang2020,Chen2022WaveYNet,Lim2022MaxwellNet,Augenstein2023}. Because of their lack of recurrence, single-step direct surrogates are substantially more time- and memory-efficient than iterative surrogates, and they have the potential to provide orders-of-magnitude speed improvements over traditional solvers like FDTD~\cite{Chen2022WaveYNet,Lim2022MaxwellNet,Augenstein2023}. These single-step solvers, however, have proven substantially more difficult to scale up to large simulation regions with many variables. This scaling difficulty is due to the large amounts of training data required to train these solvers. Prior work identifies training dataset size as a central bottleneck for robust surrogate training, with existing implementations typically scaling to a few tens of simulation variables~\cite{Jiang2020,Ma2020,Pestourie2020ActiveLearning,Pestourie2023PEDS}. Specifically, a review of neural surrogates for wave scattering~\cite{Jiang2020} observed an exponential scaling relationship between the number of variables defining the simulation (such as the number of voxels to solve for in the maps of refractive index and source values), and the number of ground truth training examples required, simulated for instance by FDTD. This same scaling is observed both when the refractive index and source distributions are sampled randomly, such as by choosing each refractive index and source pixel uniformly and independently from a predefined range, and also when training examples are pulled from a hand-curated dataset. This relationship applies when we want the trained neural surrogate to be robustly accurate, in the sense that we want the surrogate to achieve a reasonable level of accuracy across a large and diverse space of possible examples (ideally for all combinations of simulation parameters within some broad set of bounds).

By extrapolating from the dataset requirements of existing studies, a 1000-variable surrogate would likely require on the order of $10^9$ ground-truth-labeled examples under conventional training~\cite{Jiang2020}. Even if each FDTD simulation required only 0.1 seconds, generating that dataset would require approximately 3.17 years of serial compute time, as well as about 393 TB of storage assuming a $128\times128$-pixel simulation region using 64-bit complex floating point encoding. This is infeasible without a significant supercomputing effort. Additionally, these scale limitations contrast starkly with the current scale of FDTD simulations used in modern research and industry, such as in the design of freeform refractive-index devices and beam splitters, in which the number of simulation variables commonly reaches into the millions~\cite{Piggott2015Demux,Su2017,Kang2024LargeScaleInverseDesign}. 

Robust accuracy is particularly important for inverse problem applications: even if an inverse problem is initialized with a refractive index and/or source configuration which a non-robust surrogate is capable of simulating accurately, it must then sequentially traverse a diverse trajectory of examples in order to solve the desired inverse problem. If, at any point, the surrogate encounters a simulation which it is incapable of solving accurately, the refractive index and/or source gradients will be commensurately inaccurate and the inverse problem may fail to reach a satisfactory optimal. Thus, we require a neural surrogate which is sufficiently accurate across its entire domain of possible simulations such that encountering a high-loss simulation, in which a chosen loss function between the surrogate prediction and the ground truth FDTD-produced field is unacceptably high, is sufficiently unlikely. 

The poor scaling of existing training methods for single-step neural surrogates is also an issue when considering the potential of foundation-model-like neural simulators for wave scattering inverse problems, an aspiration related to broader work on neural operators and learned PDE solution maps~\cite{Li2021FNO,Lu2021DeepONet,Kovachki2023NeuralOperator,Raissi2019}. Considering that traditional simulators like FDTD are able to simulate near-arbitrary dense grids of refractive indices and sources, and inverse problems typically involve traversing through a diverse series of simulations without simple predetermined structure, any replacement will need to replicate this capability. Doing so requires many simulation variables: for instance, a square region 100 wavelengths, $\lambda$, across with grid spacing $\lambda/10$ will require about 3 million simulation variables, infeasible with current training techniques for single-step neural surrogates. 

In this work, we propose a training strategy aimed at high inference speed, robust accuracy and inductive scalability, diagrammed in Fig.~\ref{diagram}. Rather than assembling a fixed training dataset in advance, or randomly sampling each simulation variable independently, we run a parallel process that dynamically generates training examples during learning by explicitly searching for examples on which the current surrogate disagrees with the ground truth (FDTD). The key idea is to treat the training data as optimization variables and to continuously perform a constrained, gradient-based hill climb in the space of the inputs (refractive index fields and complex-valued sources) to find examples that maximize a surrogate--ground truth mismatch objective. Each hill-climbed instance is labeled by the ground-truth solver and added to a dynamic training dataset, from which the surrogate is trained continuously. As the surrogate improves, the generator automatically shifts its attention to new failure modes, producing a curriculum of increasingly informative examples without requiring exhaustive random coverage. This method is partly inspired by adversarial robustness strategies used in computer vision, in which networks are trained on examples deliberately chosen to expose model failures, typically for classification or regression tasks~\cite{Goodfellow2014,Madry2018}. However, rather than seeking robustness to imperceptible perturbations around a fixed dataset, our goal is to efficiently explore the physically valid space of scattering problems and identify examples that are maximally informative for learning the global surrogate map. In this sense, our approach combines adversarial example generation, active learning, online hard-example mining, and online dataset construction~\cite{Settles2009,Brinker2003,Shrivastava2016OHEM}: the training distribution is not fixed, but is continuously reshaped by the current model's weaknesses. We find that this dramatically improves the efficiency of training, allowing a single-step surrogate to reach robust accuracy in regimes where random sampling would require an impractically large number of labeled simulations.

We find that the careful management of high-intensity fields caused by near-perfect resonances is critical to the convergence and stability of our method, and for this reason we propose resonance management strategies using source and gradient normalization, specifically using a relative loss measure for the search and training process which emphasizes structural differences between the predicted and ground truth fields as opposed to differences in their relative magnitude. Further, by leveraging multi-scale training, the spatial locality of a U-Net-based convolutional model architecture~\cite{Ronneberger2015UNet,Long2015FCN}, and a spatially local FDTD ground truth solver, we take advantage of an intrinsic inductive bias, such that the model can provide accurate simulations for regions much larger than those used during training. In our experiments, we find that our method provides robust accuracy across a broad set of challenging simulation examples, and inductively scales (generalizes to larger domains than seen during training) to regions at least $64\times$ larger than trained on, and provides comparable or better FDTD-validated performance in freeform inverse problems, with optimization-time speedups ranging from $1.29\times$ to $26.5\times$ across the inverse-design workloads tested here.

\section{Results and Discussion}

Our goal is to train a single-step neural surrogate that remains accurate across a broad, high-dimensional family of two-dimensional wave-scattering problems. This is a significantly more challenging task than achieving low average loss on a randomly selected or hand-curated dataset; a surrogate which appears to be accurate for such a pre-defined training dataset will typically still have a large number of examples for which the surrogate is severely inaccurate, as shown in Fig.~\ref{fig:128_examples}, which shows a set of manually-curated examples exhibiting properties such as smooth variation, high-contrast resonance, and long-range structure. We find that these adversarial examples are often structured examples (e.g. ring resonators, ordered patterns of high-contrast refractive indices, etc.) involving complex, structured optical interactions. We further want our trained surrogate to scale inductively to larger simulation regions than are present in the training data, as training for larger regions is limited by computation, training time and memory. In the following experiments, we therefore evaluate the surrogate along four axes: the dataset-size efficiency of dynamic hard-example generation relative to random sampling, robustness on structured and unstructured examples outside the random training distribution, inductive scalability of the trained model, and FDTD-validated performance in downstream inverse-design tasks.

\subsection{Problem scale and surrogate form}
\label{sec:problem_scale_results}

We consider two-dimensional transverse-electric (TE) scattering on a uniform Cartesian grid. Each example is defined by a complex source field $s \in \mathbb{C}^{N\times N}$ and a refractive-index field $n \in \mathbb{R}^{N\times N}$, and the target output is the steady-state out-of-plane electric field $E_z \in \mathbb{C}^{N\times N}$. Nonperiodic boundaries use a perfectly matched layer (PML) of thickness $t$ grid cells~\cite{Berenger1994}. The PML source is fixed to zero, the PML refractive index is fixed to $n=1$, and the interior refractive index is constrained to $n\in[1,2]$ unless otherwise stated.

The key scale challenge is that both $s$ and $n$ are dense, freely-configurable fields. Counting one real-valued refractive-index variable (neglecting absorption/gain for the following discussions for simplicity) and two real-valued source variables per non-PML pixel gives $D = 3\,(N-2t)^2$ controllable simulation variables. The model is trained on $64\times64$ and $128\times128$ domains with $t=5$, corresponding to $D=8{,}748$ and $D=41{,}772$ controllable variables, respectively. In the largest evaluation below, the same trained model is applied without retraining to a $1024\times1024$ domain with $D=3{,}084{,}588$ (the number of constituent pixels minus the fixed PML boundary regions) controllable variables.

The ground truth field is computed with a GPU-accelerated FDTD solver~\cite{Yee1966,Taflove2005,Oskooi2010}, at the end of 300 timesteps. The surrogate $S_\theta$ is a fully-convolutional U-Net that maps the real and imaginary source channels and refractive index channel to the real and imaginary channels of the complex field~\cite{Ronneberger2015UNet,Long2015FCN}. The fully convolutional form is important for the inductive-scaling experiments below: no part of the architecture requires a fixed input size, so the trained network can be evaluated on grids larger than those used during training.

\subsection{Normalization for resonant examples}
\label{sec:loss_norm_rewrite}

A central difficulty in training and hard-example generation is the large variation in field energy across scattering instances. Near-resonant configurations can produce fields much larger than typical examples. If training used raw field MSE, we find that a small number of high-energy examples dominate both the surrogate update and the hard-example search.

We therefore compare predictions after normalizing each example by the RMS amplitude of its ground truth field. For a ground truth field $E$, define
\begin{equation}
r(E)=\left(\frac{1}{N^2}\sum_{x,y}|E(x,y)|^2\right)^{1/2}.
\end{equation}
The training and generator objective is the ground-truth-RMS-normalized complex-field error
\begin{equation}
\mathcal{L}(s,n;\theta)=\frac{1}{N^2}\sum_{x,y}\left|\frac{S_\theta(s,n)(x,y)-T(s,n)(x,y)}{r(T(s,n))}\right|^2.
\label{eq:teacher_rms_loss_rewrite}
\end{equation}
This loss measures relative complex-field error and places low- and high-energy examples on a comparable scale. 

\subsection{Efficient training data optimization by dynamic hard-example generation}
\label{sec:hard_example_generation_rewrite}

Figure~\ref{diagram} summarizes our dynamic training approach, which occurs in parallel to the neural surrogate model training via a `hard-example generator'. We dynamically update our training data to expose failures of the current surrogate, analogous in spirit to adversarial training, active learning, and online hard-example mining~\cite{Goodfellow2014,Madry2018,Settles2009,Shrivastava2016OHEM}. For fixed network weights $\theta$ (the current neural surrogate), the generator searches over physically-valid sources and refractive-index fields to find the ones that result in the largest loss value (difference between the surrogate's output electric field and the ground-truth (FDTD) simulated one):
\begin{equation}
(s^\star,n^\star) \in \arg\max_{(s,n)\in\mathcal{C}} \mathcal{L}(s,n;\theta),
\end{equation}
where $\mathcal{C}$ enforces the index bounds, PML constraints, and a unit-energy source constraint. The source normalization prevents the generator from increasing the loss by trivially scaling the source amplitude. When a high-loss example is discovered, its inputs (source, refractive index) and the ground-truth output are added to the training dataset, which the neural surrogate is continuously training on. The hard-example generation process periodically updates the neural surrogate model to be the most recent version. 

To keep the number of loss-maximization steps spent traversing from low-loss initializations to high-loss example regions low, we maintain a consistent set of candidate simulation configurations between loss-maximization steps, meaning we start each step initialized at the latest high-loss examples in our dataset. Because of this, each generator update takes a projected ascent step on a set of previously identified high-loss examples, with periodic re-initialization of a small fraction of candidates to maintain exploration. The generator evaluates the current mismatch against the FDTD ground truth, appends the resulting hard examples to a dynamic dataset, which takes the form of a fixed-size buffer, and the trainer randomly samples from that buffer to update the surrogate. When the buffer is filled, the oldest examples in the buffer are discarded. The use of this buffer is important for convergence, paralleling the stabilizing role of replay buffers in sequential learning systems~\cite{Lin1992ExperienceReplay,Schaul2016PrioritizedReplay}: we find that, if the data generation and model training are tightly coupled, with hard examples being used to train the model immediately after they are generated, the model will typically fail to converge. This is likely due to the adversarial nature of the process: if the newly generated examples are used immediately for training, the tightly coupled generator and trainer form an unstable feedback loop. The generator continually modifies the input distribution to increase the current surrogate error, while the trainer attempts to reduce that error using only the most recently generated examples. Because this distribution changes rapidly and consecutive examples are strongly correlated, the trainer does not receive a sufficiently broad or stable learning signal from which to improve the underlying scattering approximation. In practice, the model fails to converge and retains high loss across both newly generated and previously encountered examples. The replay buffer mitigates this instability by mixing recent hard examples with examples generated at earlier stages of training. This reduces temporal correlation between successive updates, slows the rate at which the effective training distribution changes, and provides a more representative sample of the failure modes accumulated over training, allowing the surrogate and generator to converge toward progressively more difficult examples.

\begin{figure}[]
\centering\includegraphics[width=16cm]{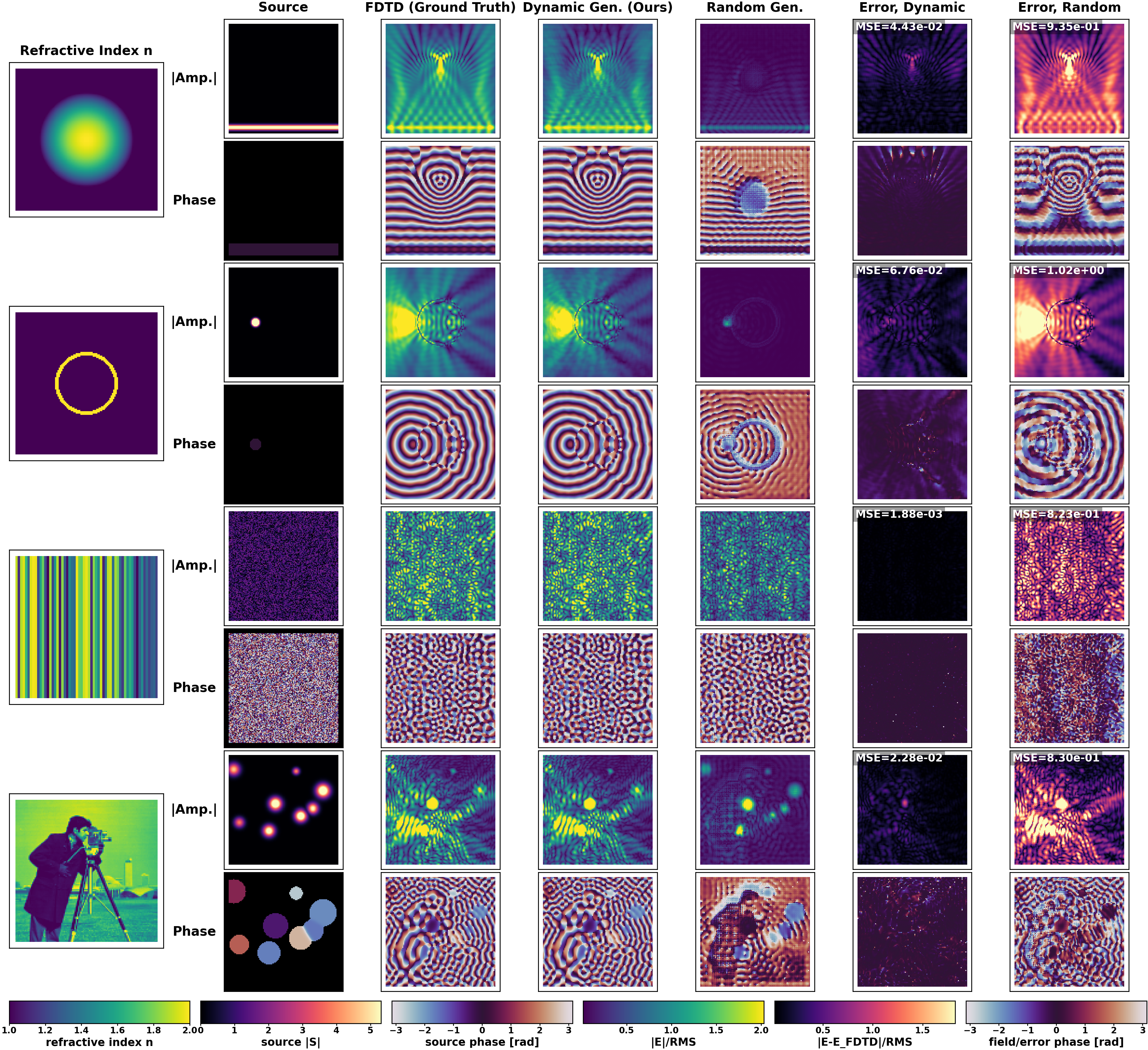}
\caption{Dynamically generating hard example training data improves neural surrogate performance. Here we show a diverse collection of challenging examples of source and refractive index configurations. We simulate the electric field ground truth with FDTD, then compare with the neural surrogate output either trained using our dynamic hard-example generation approach or traditional random sampling. Each example shows the refractive index map, as well as the magnitude and phase of the source, ground truth (FDTD) field, surrogate predictions, and error. To make maps more comparable across examples with different total intensities, error values report  complex-field error normalized by the root-mean-square (RMS) intensity of the ground truth field.}
\label{fig:128_examples}
\end{figure}

The training run used in the following experiments alternated between $64\times64$ and $128\times128$ grids. It used generator and trainer batch sizes of 50, reached 165{,}579 generator iterations and 200,000 trainer steps, and appended $8.28\times10^6$ teacher-labeled examples to the dynamic dataset. The rolling dynamic dataset retained 100{,}000 examples at a time, split evenly across the two training resolutions. Compared with random sampling, dynamic generation produced lower validation error on a validation set of challenging structured and unstructured examples in Fig.~\ref{fig:dynamic_vs_random_loss}, providing simulations with much greater physical fidelity (Fig.~\ref{fig:128_examples}).

\begin{figure}[]
\centering\includegraphics[width=12cm]{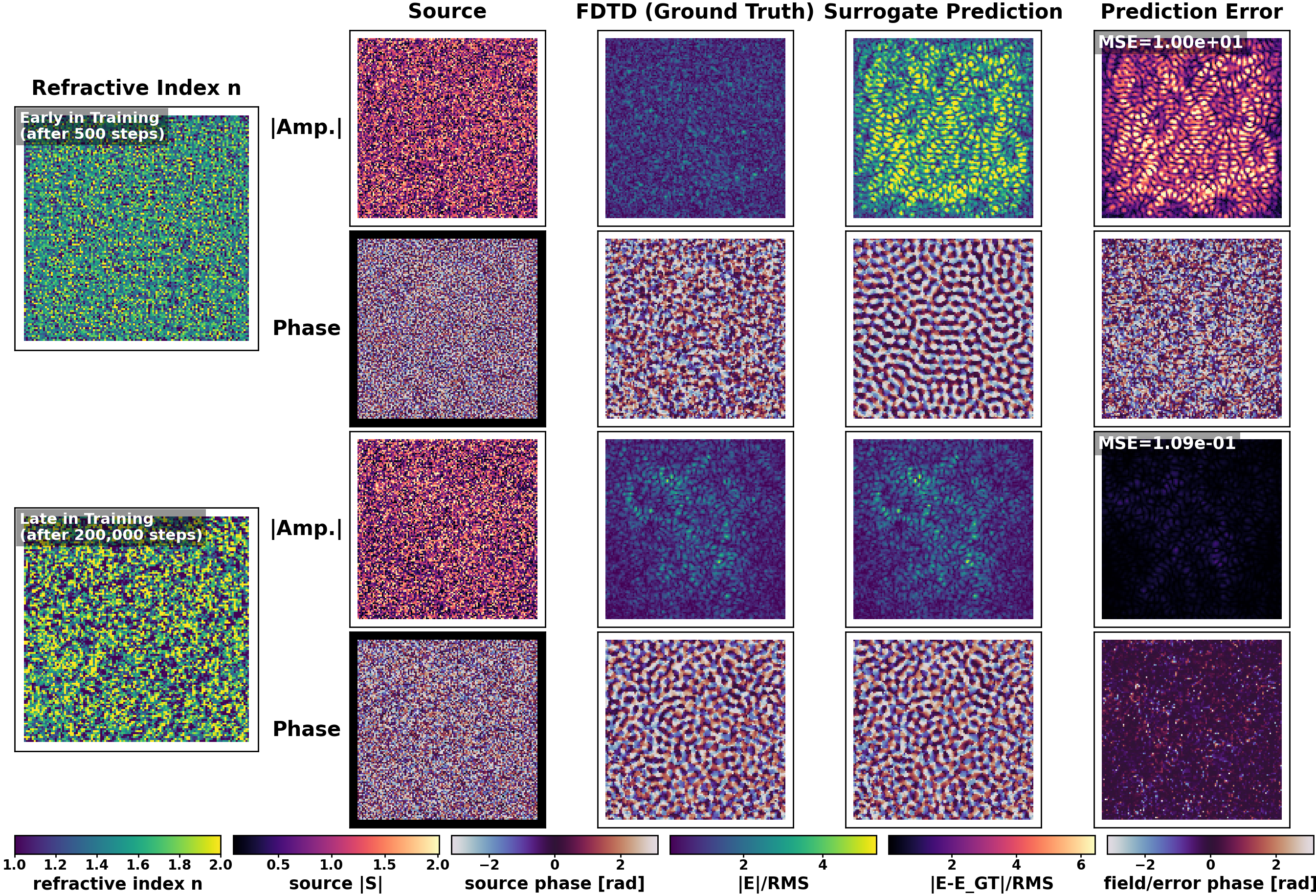}
\caption{High-loss examples generated by our dynamic dataset generation process. (Top) An example from early in training, after 500 training steps. (Bottom) An example from late in training, after 200,000 training steps. High-loss examples are rarer, become more structurally complex, and exhibit higher refractive index contrasts as training progresses, and the surrogate's error on the identified high-loss examples diminishes as it becomes more robustly accurate over the course of training.}
\label{fig:generated_hard_examples}
\end{figure}

The types of hard examples found by our dynamic generator vary over the course of the training. Figure~\ref{fig:generated_hard_examples} shows representative hard examples found early and late in training. Early in training, the generator identifies high-loss examples for which the surrogate makes large field-prediction errors. Later in training, as the neural surrogate becomes more refined, the generated examples produce less mismatch between the neural surrogate and the ground-truth FDTD. Later hard examples are qualitatively more structurally complex. Hence, we can conclude that the dynamic dataset continues to expose new failure modes as the model improves without relying on a fixed hand-designed training distribution.

\subsection{Robust accuracy on structured and unstructured inverse problems}
\label{sec:structured_accuracy_results}

We tested the trained model on a manually-curated suite of examples chosen to probe failure modes that are poorly represented by independent random sampling; some examples are shown in Fig.~\ref{fig:128_examples}. The suite includes localized and multi-source excitation, waveguide-like and ring-like structures, graded-index media, periodic grating-like index structures, randomly-placed scatterers, dense random binary media, random stripe media, and image-derived refractive-index distributions. Our dynamically trained model recovers both the amplitude and phase of the electric field across all examples, with the largest visible errors occurring near high-contrast interfaces, low-amplitude phase regions, and fine interference features. In contrast, the typical random-sampling approach, shown for comparison, leaves large structured errors across all examples. These results demonstrate the major improvements provided by our dynamic sampling approach compared to random sampling.

\subsection{Inductive scaling to larger spatial domains}
\label{sec:inductive_scaling_results}

A direct surrogate trained on one grid size does not typically generalize to larger spatial domains. Here, the fully-convolutional U-Net, the locality of the FDTD ground truth, and the multiscale training distribution together provide an inductive bias toward size generalization~\cite{Ronneberger2015UNet,Long2015FCN}. 

This size generalization is supported by two related properties of the scattering problem. First, Maxwell’s equations are local in space and time: the field at a given location is governed by nearby material properties and fields, and information propagates at a finite speed. Because our FDTD ground truth simulator is evaluated after a fixed number of time steps, each output pixel depends only on inputs within a finite causal neighborhood. The surrogate’s convolutional, multiscale architecture is well matched to this structure, as it learns local update relationships while using downsampling paths to capture interactions over progressively larger spatial scales. Second, away from the PML boundaries, the governing equations are translation equivariant: translating a source and refractive-index configuration produces a corresponding translation of the resulting field. Convolutional filters encode the same translation-equivariant bias, allowing scattering relationships learned at one location or on one domain size to be reused elsewhere and on larger domains. Training on multiple grid sizes further encourages the network to learn these spatially reusable relationships rather than features tied to a particular input dimension. Together, the space-time locality of the finite-duration scattering operator and its approximate translation equivariance provide a physical and architectural basis for inductive scaling, although they do not guarantee accuracy for propagation distances, resonances, or boundary interactions outside those represented during training.

We tested this by evaluating the same trained checkpoint on structured examples at progressively larger grid sizes while keeping the physical discretization relative to wavelength fixed.

\begin{figure}[]
\centering\includegraphics[width=12cm]{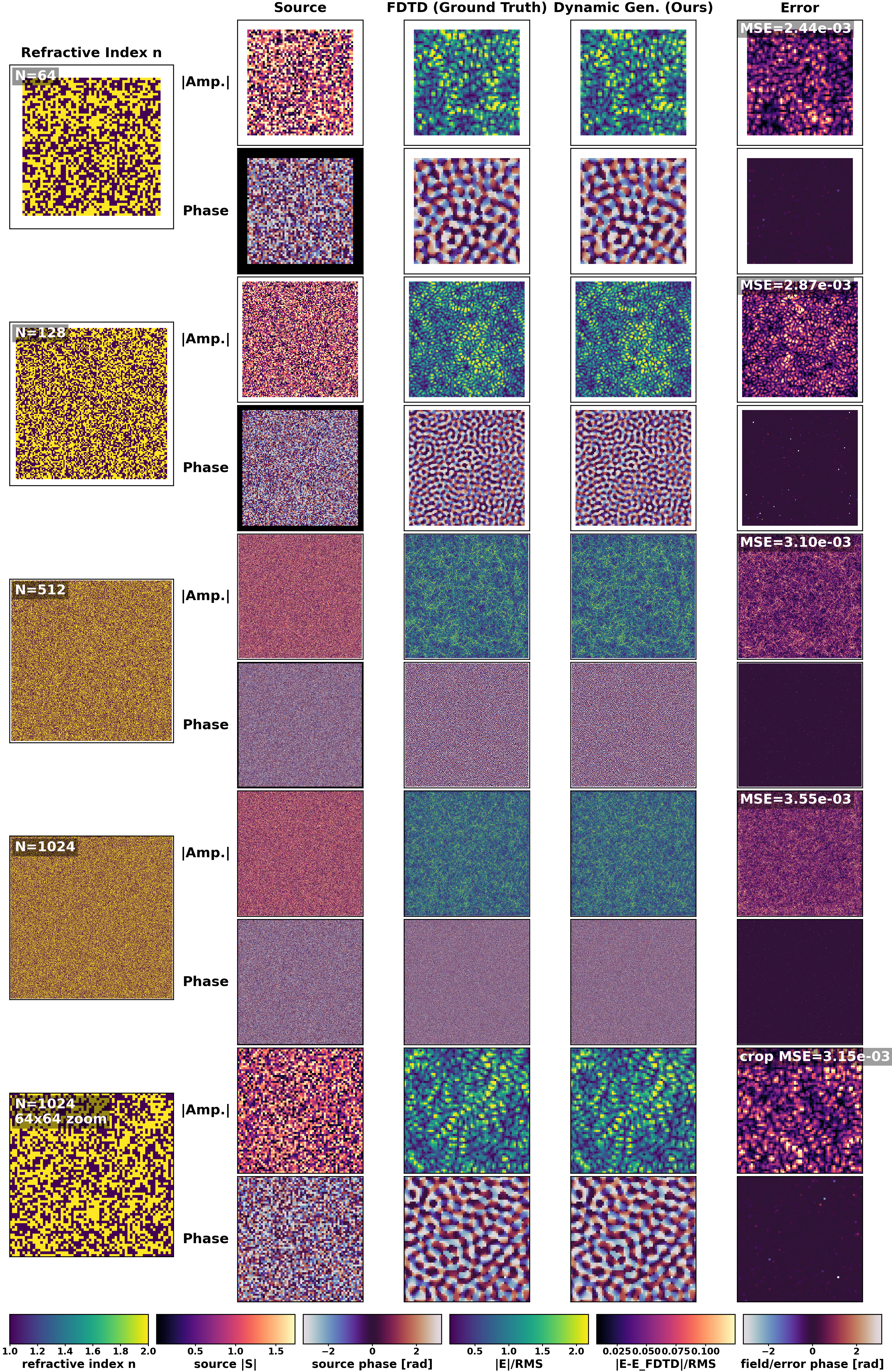}
\caption{Our method inductively scales to simulation sizes much larger than it was trained on. Here, we show an example simulation of a high-contrast binary structure at increasing grid sizes, covering a progressively larger area. For each scale, we show amplitude and phase of the refractive index/source inputs, the FDTD ground truth field, our neural surrogate prediction, and the normalized prediction error. MSE values report ground-truth-RMS-normalized complex-field error.}
\label{fig:inductive_scaling}
\end{figure}

As shown in Fig.~\ref{fig:inductive_scaling}, and further demonstrated in the following GRIN-lens inverse-design experiments, the model was trained on $64\times64$ and $128\times128$ domains (with multiple domain sizes used to encourage generalization across scales) but was evaluated without retraining on up to $1024\times1024$. The largest example corresponds to an approximately $98.1\lambda$ non-PML interior width and 3{,}084{,}588 controllable variables. This is a $73.8\times$ increase in controllable variables relative to the largest training problem and a $64\times$ increase in total grid area. The MSE values for each scale are reported in Fig.~\ref{fig:inductive_scaling}, holding consistent across scales with a very small increase with scale. We stopped at this scale purely due to memory constraints, but a more memory-optimized model using, for instance, eight-bit mixed float precision, could likely scale to much larger regions. It should be noted that the same number of FDTD cycles is predicted for each scale, so the FDTD and surrogate simulations may not capture some long-distance interactions for very large simulations unless sufficient FDTD cycles are used during training. We have not found this to be an issue for inverse problems like large-aperture GRIN-lens design, in which the strongest interactions are typically somewhat local regardless. 

These results suggest that the model has learned a local multiscale scattering representation that can be applied convolutionally to domains well beyond the training sizes. This inductive scaling is central to the practical value of the surrogate. Training directly on every target device scale would erode much of the computational advantage of neural simulation. By contrast, our model, when trained on modestly sized domains, can be deployed on substantially larger design problems, as demonstrated in the inverse-design experiments below.

\subsection{Inverse design of freeform GRIN lenses}
\label{sec:inverse_grin_results}

We then tested whether the surrogate is accurate enough to provide useful gradients for iterative inverse design, a central use case in topology-optimized photonics and freeform refractive-index optics~\cite{Molesky2018,LalauKeraly2013,Fan2020,Kang2024LargeScaleInverseDesign}. We optimized each design either through the neural surrogate or the differentiable FDTD ground-truth solver, and then validated all final designs with the FDTD ground truth.

The first task is a freeform GRIN lens illuminated by a normally incident plane wave. The design variable is a continuous refractive-index distribution inside a finite lens region, parameterized by a projected density variable. We considered two design problems: a $512\times512$ simulation of an approximately $49\lambda$-wide focusing GRIN lens and a $1024\times1024$ simulation of an approximately $98\lambda$-wide GRIN lens array.

\begin{figure}[]
\centering\includegraphics[width=14cm]{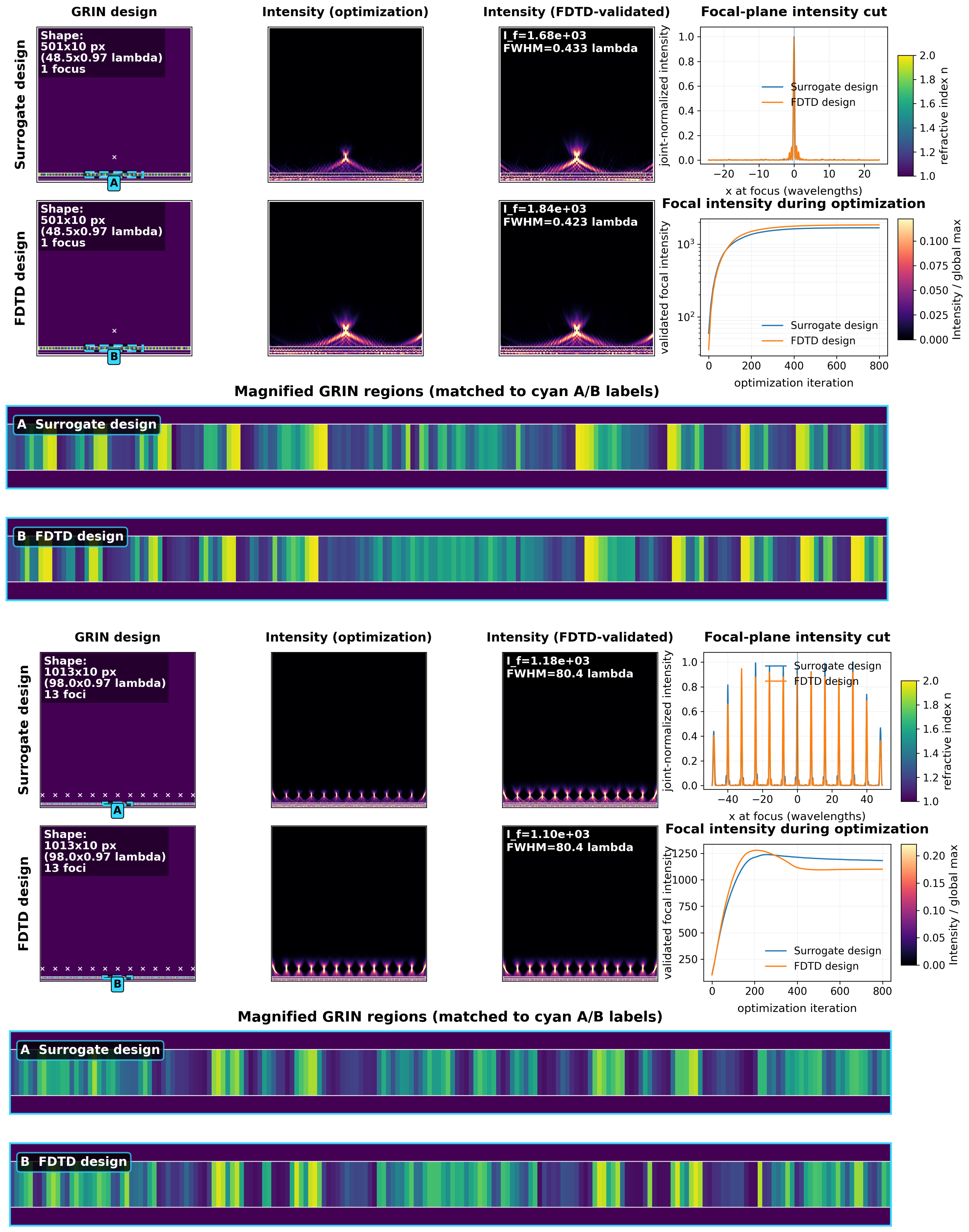}
\caption{Inverse design of freeform GRIN lenses, demonstrating accuracy and inductive scalability of surrogate. Top: a $512\times512$ simulation designs an approximately $49\lambda$-wide gradient-index (GRIN) lens that focuses a plane wave to a single point. Bottom: a $1024\times1024$ simulation designs an approximately $98\lambda$-wide GRIN lens that splits the plane wave into a periodic array of focal spots. For each case, surrogate- and FDTD-optimized designs are compared by refractive index, predicted E-field at the end of optimization (from the neural surrogate model or FDTD respectively), the FDTD-validated E-field of the finished designs, focal-plane cut, optimization history, and matched zoomed sections of the GRIN lenses.}
\label{fig:inverse_grin_lens}
\end{figure}

Figure~\ref{fig:inverse_grin_lens} shows that the surrogate-optimized GRIN lenses transfer to the FDTD ground truth. For the $48.5\lambda$ focusing design, the NN-optimized structure achieved a validated focal intensity of 1669, compared with 1839 for the FDTD-optimized design. The corresponding focal-plane FWHM values were $0.434\lambda$ and $0.423\lambda$. The NN optimization required 92.2s compared with 474s for direct FDTD optimization, giving a $5.14\times$ speedup.

For the $98.0\lambda$ periodic-focus design, the surerogate-optimized structure achieved a mean target intensity of 1221 across 13 focal points, compared with 1147 for the FDTD-optimized design. The target-to-target coefficients of variation were 0.222 and 0.246, respectively, indicating slightly more uniform target intensities for the surrogate-optimized design. Surrogate-based optimization required 372s compared with 479s for FDTD, corresponding to a $1.29\times$ speedup. Thus, the surrogate produced high-performing FDTD-validated GRIN designs at both tested scales while reducing wall-clock optimization time.

The NN and FDTD optimizers converged to similar structures while producing comparable validated fields, with the NN performing slightly better than FDTD for the larger $98.0\lambda$ design. We believe this improvement may be due to a regularizing effect with the NN. The large-area, many-point loss function is likely somewhat "spiky," and difficult to optimize, and the NN appears to have an easier time maintaining a high-quality design over many iterations without jumping into a less-functional local optimum. We found this effect to be preserved over a range of hyperparameter settings, such as different learning rates.
 
\subsection{Inverse design of a waveguide beam splitter}
\label{sec:inverse_beamsplitter_results}

The second inverse-design task is a compact freeform waveguide junction, a device class closely related to prior integrated-photonic inverse-design demonstrations~\cite{Lu2013NanophotonicComputationalDesign,Piggott2015Demux,Su2017}. A guided mode input enters a square design region on the left, and the objective is to route equal guided-mode-coupled power into top and bottom output waveguides. Unlike a focal-intensity objective, this task evaluates modal overlap with the forward-propagating fundamental mode of each output waveguide.

\begin{figure}[]
\centering\includegraphics[width=14cm]{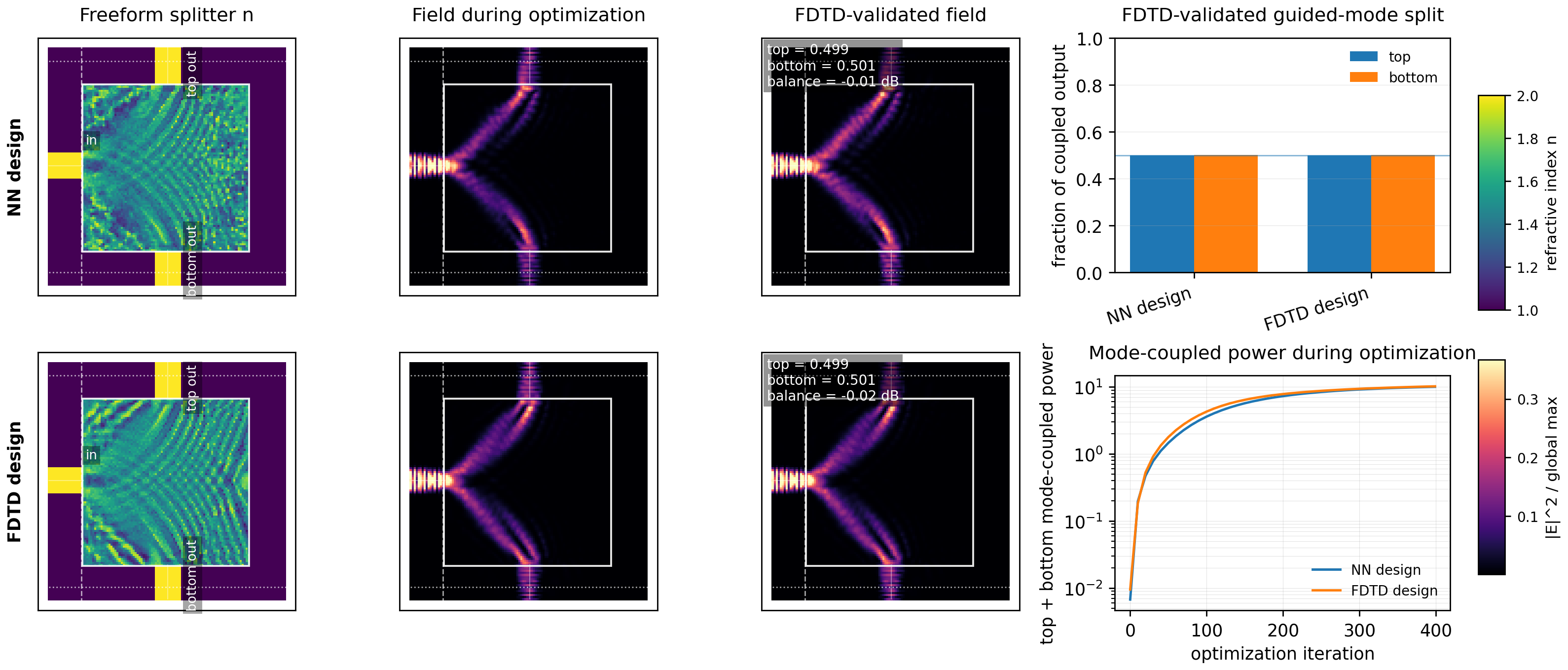}
\caption{Inverse design of a waveguide beam splitter. A $128\times128$ simulation designs an $8\lambda \times 8\lambda$ freeform junction that routes a left-input waveguide mode equally into top and bottom output waveguides. Surrogate- and FDTD-optimized designs are compared by refractive index, optimization field, FDTD-validated field, guided-mode output split, and mode-coupled power during optimization.}
\label{fig:inverse_beamsplitter}
\end{figure}

As shown in Fig.~\ref{fig:inverse_beamsplitter}, the surrogate-optimized design produced (with each final design validate by the FDTD solver) top and bottom guided-mode powers of 5.41 and 5.15, respectively, corresponding to a $51.2/48.8\%$ split and a top-over-bottom balance of $+0.213$ dB. The direct FDTD-optimized design produced top and bottom guided-mode powers of 5.11 and 5.13, corresponding to a $49.9/50.1\%$ split and a balance of $-0.018$ dB. The surrogate design therefore achieved a comparable split balance and a higher total guided-mode-coupled output power, 10.55 compared with 10.24 for direct FDTD optimization, with lower total leakage, 0.137 compared with 0.152. Excluding final FDTD validation/evaluation time, the surrogate optimization required 8.97s compared with 238.0s for the FDTD baseline, giving a $26.5\times$ optimization-loop speedup. This result indicates that the surrogate can provide useful gradients for photonic inverse-design objectives, producing FDTD-validated designs with comparable or better total guided output power while substantially reducing optimization time.

\begin{table*}[t]
\centering
\caption{FDTD-validated inverse-design performance and optimization timing. Optimization time includes only the recorded optimization loop and excludes final FDTD validation/evaluation, figure writing, model compilation, warmup, and other script overhead. Speedup is defined as $t_{\mathrm{FDTD}}/t_{\mathrm{NN}}$.}
\label{tab:inverse_timing}
\footnotesize
\setlength{\tabcolsep}{4pt}
\renewcommand{\arraystretch}{1.12}
\begin{tabular*}{\textwidth}{@{\extracolsep{\fill}}p{2.6cm}ccrrrp{4.0cm}@{}}
\hline
Task & Grid & Steps & \multicolumn{1}{c}{NN (s)} & \multicolumn{1}{c}{FDTD (s)} & Speedup & FDTD-validated result, NN / FDTD \\
\hline
Single-focus GRIN lens & $512^2$ & 800 & 92.2 & 474 & $5.14\times$ & Focal intensity: $1.67{\times}10^3$ / $1.84{\times}10^3$ \\
Periodic-focus GRIN lens & $1024^2$ & 800 & 372 & 479 & $1.29\times$ & Mean target intensity: $1.22{\times}10^3$ / $1.15{\times}10^3$ \\
50:50 beam splitter & $128^2$ & 400 & 8.97 & 238 & $26.5\times$ & Total mode power: 10.55 / 10.24; Split: $51.2/48.8\%$ / $49.9/50.1\%$ \\
\hline
\end{tabular*}
\end{table*}

\subsection{Limitations and scope}
\label{sec:limitations_results}

These results demonstrate robust accuracy and useful inverse-design performance for a single-wavelength, two-dimensional transverse-electric setting with bounded refractive-index contrast. The current model does not yet address full vectorial three-dimensional scattering, broadband response, material dispersion, fabrication constraints, or arbitrary boundary conditions. Although the model generalizes to spatial domains much larger than those used during training, it should not be expected to remain accurate under substantially different refractive-index ranges, grid resolutions relative to wavelength, boundary conventions, or source parameterizations without additional training. Finally, dynamic hard-example generation improves empirical coverage of difficult cases but does not provide a formal worst-case error bound.

Nonetheless, we believe that the training method described here is the first to bypass major previous limitations on the scaling, speed, and adoption of neural surrogates for wave simulation. Dynamic hard-example generation shifts data collection away from exhaustive random coverage and toward the surrogate's current weaknesses. When combined with normalization, replay buffering, and a fully convolutional architecture, this strategy produces a reusable single-step surrogate that predicts fields across challenging scattering examples, scales inductively to larger domains, and provides gradients that transfer to FDTD-validated inverse designs. While, due to resource availability, the surrogate in this study was trained only on two GPUs, future work could readily apply the described algorithm across much larger compute clusters, potentially distributing the ground truth simulation and surrogate training process across hundreds of GPUs. In these cases, we anticipate that our algorithm will be able to train fast, robustly accurate neural surrogates for wave scattering with much larger spatial scales: potentially 100,000s of wavelengths across with currently available hardware. Future work may also extend our training strategy to broadband, vectorial, and three-dimensional electromagnetics, potentially with the incorporation of formal error bounds for verifiable accuracy during operation. The training strategies described here may also be of interest for the training of surrogate models in other areas of physics, such as for heat propagation and fluid flow.

\section{Methods}

\subsection{Training data generation}

Training examples are generated online. At each generator iteration, a batch of source and refractive-index fields is constructed on a user-specified square grid. The generator alternated between $64\times64$ and $128\times128$ domains, used a batch size of 50, and used one Adam ascent step per generator update \cite{Kingma2015Adam}. Candidate configurations were preserved between generator steps, with a small probability for each example to be independently reinitialized after each generator step, chosen as $10^{-3}$. Reinitialized candidates were sampled uniformly randomly on a pixel-by-pixel basis. All examples are projected back to the feasible set after every update: $n$ is clipped to the allowed range $[1,2]$, the PML ring of thickness 5 pixels is fixed to $n=1$, the source is zeroed inside the PML, and the complex source is projected to unit energy.

The training buffer stores the source, refractive-index, and teacher fields. By the final logged point, 8.28 million examples had been appended to the replay stream and 8.18 million had been dropped by the FIFO replacement rule; the active FIFO held 100,000 examples. This replay-buffer mechanism lets the dynamic dataset change and adapt through the course of training while reducing the instability associated with immediately training only on the newest hard examples \cite{Lin1992ExperienceReplay,Schaul2016PrioritizedReplay}.

\subsection{Neural network model}

The surrogate is a fully convolutional U-Net-style encoder--decoder with skip connections \cite{Ronneberger2015UNet,Long2015FCN}. Each input example is represented as a tensor in $\mathbb{R}^{3\times N\times N}$ containing $\mathrm{Re}(s)$, $\mathrm{Im}(s)$, and the refractive-index field $n$. The network outputs a tensor in $\mathbb{R}^{2\times N\times N}$, interpreted as $\mathrm{Re}(\hat{E}_z)$ and $\mathrm{Im}(\hat{E}_z)$.

The model begins with a $3\times3$ convolution from 3 channels to 128 channels. The encoder has six downsampling stages with channel widths
\[
128,\;256,\;512,\;512,\;512,\;512,\;512,
\]
where the maximum channel width is capped at 512. At each encoder resolution, the model applies four residual convolutional blocks before downsampling. Each residual block consists of two $3\times3$ convolutions, each followed by group normalization and a SiLU nonlinearity, with a residual skip connection and a $1\times1$ projection when the number of channels changes. Downsampling is performed by a stride-2 $3\times3$ convolution. The bottleneck contains five residual convolutional blocks at the coarsest resolution.

The decoder mirrors the encoder. Each upsampling stage uses a $3\times3$ convolution followed by $2\times$ pixel shuffle upsampling, concatenates the corresponding encoder skip feature map, and then applies four residual convolutional blocks. A final $1\times1$ convolution maps the 128-channel feature map to the two output field channels. All convolutions use nonperiodic zero padding, matching the PML/open-boundary setting rather than a periodic domain.

The trained surrogate used in the reported experiments contains 253{,}658{,}118 trainable parameters in the field-prediction network. Because the architecture is fully convolutional and contains no dense layers tied to a particular spatial grid size, the same learned weights can be evaluated on larger grids than those used during training. This size transfer was validated empirically in the preceding inductive-scaling experiments.

\subsection{Surrogate training}

The trainer samples minibatches from the replay buffer and minimizes the teacher-RMS-normalized complex-field loss in Eq.~\ref{eq:teacher_rms_loss_rewrite}. The model is optimized with Adam using learning rate $10^{-4}$, batch size 50, no weight decay, and no gradient clipping \cite{Kingma2015Adam}. All training runs and ablation studies were run for 200,000 trainer optimizer steps. To increase training speed, we run the generator and trainer in parallel on separate GPUs; for this run we used two NVIDIA RTX Pro 6000 Blackwell, with 96 GB of memory each. The trainer synchronized its weights to the generator every trainer step.

For the random-sampling baseline, the same surrogate architecture and loss are used, but the training data are drawn from the base random distribution rather than from gradient-based hard-example generation. This isolates the effect of the data-generation strategy from the model architecture.

\subsection{Structured validation examples}

The structured validation set is manually curated and partly procedurally generated. It includes simple point-source and two-source examples, a high-index block, waveguide-like structures, a ring-like resonator, a graded-index lens, a periodic grating-like index structure, randomly placed scatterers, dense random binary media, random stripe media, and an image-derived refractive-index field with random source spots. For each example, the FDTD teacher is run using the same boundary and discretization settings as training. The reported MSE values are computed after dividing the prediction error by the teacher RMS, consistent with Eq.~\ref{eq:teacher_rms_loss_rewrite}.

\subsection{Inductive-scaling evaluation}

For inductive scaling, the same trained checkpoint is evaluated on a fixed structured example at multiple grid sizes. The physical discretization relative to wavelength is held fixed, and the example construction is repeated at each scale, such that the spatial size of the region scales with the grid size. The fully convolutional surrogate is applied directly to the larger grids without retraining or architectural modification \cite{Long2015FCN,Ronneberger2015UNet}. The teacher field is computed independently for each scale, and the reported errors are teacher-RMS-normalized MSE and relative $L_2$ error.

\subsection{Freeform GRIN-lens inverse design}

The GRIN-lens design variable is a projected density field $\rho\in[0,1]$ defined inside a finite lens region, following the general density-based inverse-design and topology-optimization paradigm used in photonic design \cite{Molesky2018,Fan2020,Kang2024LargeScaleInverseDesign}. It is mapped to refractive index as
\begin{equation}
n(x,y)=n_{\mathrm{low}}+(n_{\mathrm{high}}-n_{\mathrm{low}})\rho(x,y)
\end{equation}
inside the design region and to the background index outside. The incident plane wave is represented by a narrow horizontal complex source band. The optimization objective maximizes the intensity at the target focal point, with the multi-point objective taking the average intensity of all focal points. Surrogate designs are optimized by backpropagation through the neural model; FDTD baselines are optimized using the differentiable teacher. The reported GRIN timing runs use compiled surrogate evaluation with bfloat16 autocast for the neural optimization path, while all final designs are validated with the FDTD teacher.

\subsection{Beam-splitter inverse design}

The beam-splitter design variable is again a projected density field inside a square freeform junction, in the spirit of compact integrated-photonic inverse-designed devices \cite{Lu2013NanophotonicComputationalDesign,Piggott2015Demux,Su2017}. A left-input waveguide and top/bottom output waveguides are fixed high-index regions, while the junction is optimized. Output powers are measured by overlap with the forward-propagating fundamental mode of each output waveguide. The objective maximizes total top-plus-bottom guided output power and penalizes deviation from a 50:50 split,
\begin{equation}
\mathcal{L}_{\mathrm{split}}=-\log(P_{\mathrm{top}}+P_{\mathrm{bottom}}),
\end{equation}
As with the GRIN-lens experiment, the NN and FDTD optimization paths are both validated using the FDTD teacher.

\section{Conclusions}

We demonstrated a single-step neural surrogate for dense, freeform two-dimensional wave-scattering problems with thousands to millions of configurable source and material variables. The main obstacle to this regime is data generation: random sampling wastes teacher simulations on easy examples and leaves systematic failure modes that can be exploited by downstream optimizers. By dynamically generating hard examples through gradient ascent on the surrogate--teacher mismatch, and by stabilizing the process with teacher-RMS normalization, source normalization, and replay buffering, we train a surrogate that is robust across a diverse set of scattering configurations.

The resulting model accurately predicts complex FDTD fields for challenging structured and unstructured examples, generalizes inductively to domains much larger than those used during training, and provides useful gradients for inverse design. In freeform GRIN-lens and waveguide beam-splitter design tasks, optimization through the neural surrogate yields FDTD-validated devices with performance comparable to direct FDTD optimization, reducing optimization-loop time by $5.14\times$ for the $48.5\lambda$ single-focus GRIN lens, $1.29\times$ for the $98.0\lambda$ periodic-focus GRIN lens, and $26.5\times$ for the compact beam splitter. These results suggest a practical route toward a foundation model for the neural simulation of wave scattering, one capable of achieving faster inference than traditional simulators, robust accuracy across diverse simulation problems, and inductive scaling up to scales useful for problems of current interest in large-scale photonic inverse design.



\printbibliography

@article{Yee1966,
  author = {Yee, Kane S.},
  title = {Numerical Solution of Initial Boundary Value Problems Involving {Maxwell}'s Equations in Isotropic Media},
  journal = {IEEE Transactions on Antennas and Propagation},
  volume = {14},
  number = {3},
  pages = {302--307},
  year = {1966},
  doi = {10.1109/TAP.1966.1138693}
}

@article{Berenger1994,
  author = {Berenger, Jean-Pierre},
  title = {A Perfectly Matched Layer for the Absorption of Electromagnetic Waves},
  journal = {Journal of Computational Physics},
  volume = {114},
  number = {2},
  pages = {185--200},
  year = {1994},
  doi = {10.1006/jcph.1994.1159}
}

@book{Taflove2005,
  author = {Taflove, Allen and Hagness, Susan C.},
  title = {Computational Electrodynamics: The Finite-Difference Time-Domain Method},
  publisher = {Artech House},
  address = {Norwood, MA},
  edition = {3rd},
  year = {2005}
}

@article{Oskooi2010,
  author = {Oskooi, Ardavan F. and Roundy, David and Ibanescu, Mihai and Bermel, Peter and Joannopoulos, J. D. and Johnson, Steven G.},
  title = {{Meep}: A Flexible Free-Software Package for Electromagnetic Simulations by the {FDTD} Method},
  journal = {Computer Physics Communications},
  volume = {181},
  number = {3},
  pages = {687--702},
  year = {2010},
  doi = {10.1016/j.cpc.2009.11.008}
}

@article{Molesky2018,
  author = {Molesky, Sean and Lin, Zin and Piggott, Alexander Y. and Jin, Weiliang and Vuckovi{\'c}, Jelena and Rodriguez, Alejandro W.},
  title = {Inverse Design in Nanophotonics},
  journal = {Nature Photonics},
  volume = {12},
  number = {11},
  pages = {659--670},
  year = {2018},
  doi = {10.1038/s41566-018-0246-9}
}

@article{LalauKeraly2013,
  author = {Lalau-Keraly, Christopher M. and Bhargava, Samarth and Miller, Owen D. and Yablonovitch, Eli},
  title = {Adjoint Shape Optimization Applied to Electromagnetic Design},
  journal = {Optics Express},
  volume = {21},
  number = {18},
  pages = {21693--21701},
  year = {2013},
  doi = {10.1364/OE.21.021693}
}

@article{Hughes2019,
  author = {Hughes, Tyler W. and Williamson, Ian A. D. and Minkov, Momchil and Fan, Shanhui},
  title = {Forward-Mode Differentiation of {Maxwell}'s Equations},
  journal = {ACS Photonics},
  volume = {6},
  number = {11},
  pages = {3010--3016},
  year = {2019},
  doi = {10.1021/acsphotonics.9b01238}
}

@article{Minkov2020,
  author = {Minkov, Momchil and Williamson, Ian A. D. and Andreani, Lucio C. and Gerace, Dario and Lou, Beicheng and Song, Alex Y. and Hughes, Tyler W. and Fan, Shanhui},
  title = {Inverse Design of Photonic Crystals through Automatic Differentiation},
  journal = {ACS Photonics},
  volume = {7},
  number = {7},
  pages = {1729--1741},
  year = {2020},
  doi = {10.1021/acsphotonics.0c00327}
}

@article{Lu2013NanophotonicComputationalDesign,
  author = {Lu, Jesse and Vuckovi{\'c}, Jelena},
  title = {Nanophotonic Computational Design},
  journal = {Optics Express},
  volume = {21},
  number = {11},
  pages = {13351--13367},
  year = {2013},
  doi = {10.1364/OE.21.013351}
}

@article{Piggott2015Demux,
  author = {Piggott, Alexander Y. and Lu, Jesse and Lagoudakis, Konstantinos G. and Petykiewicz, Jan and Babinec, Thomas M. and Vuckovi{\'c}, Jelena},
  title = {Inverse Design and Demonstration of a Compact and Broadband On-Chip Wavelength Demultiplexer},
  journal = {Nature Photonics},
  volume = {9},
  number = {6},
  pages = {374--377},
  year = {2015},
  doi = {10.1038/nphoton.2015.69}
}

@article{Su2017,
  author = {Su, Logan and Piggott, Alexander Y. and Sapra, Neil V. and Petykiewicz, Jan and Vuckovi{\'c}, Jelena},
  title = {Inverse Design and Demonstration of a Compact On-Chip Narrowband Three-Channel Wavelength Demultiplexer},
  journal = {ACS Photonics},
  volume = {5},
  number = {2},
  pages = {301--305},
  year = {2018},
  doi = {10.1021/acsphotonics.7b00987}
}

@article{Kang2024LargeScaleInverseDesign,
  author = {Kang, Chanik and Park, Chaejin and Lee, Myunghoo and Kang, Joonho and Jang, Min Seok and Chung, Haejun},
  title = {Large-Scale Photonic Inverse Design: Computational Challenges and Breakthroughs},
  journal = {Nanophotonics},
  volume = {13},
  number = {20},
  pages = {3765--3792},
  year = {2024},
  doi = {10.1515/nanoph-2024-0127}
}

@article{Fan2020,
  author = {Fan, Jonathan A.},
  title = {Freeform Metasurface Design Based on Topology Optimization},
  journal = {MRS Bulletin},
  volume = {45},
  number = {3},
  pages = {196--201},
  year = {2020},
  doi = {10.1557/mrs.2020.62}
}

@article{Jiang2020,
  author = {Jiang, Jiaqi and Chen, Mingkun and Fan, Jonathan A.},
  title = {Deep Neural Networks for the Evaluation and Design of Photonic Devices},
  journal = {Nature Reviews Materials},
  volume = {6},
  number = {8},
  pages = {679--700},
  year = {2021},
  doi = {10.1038/s41578-020-00260-1}
}

@article{Ma2020,
  author = {Ma, Wei and Liu, Zhaocheng and Kudyshev, Zhaxylyk A. and Boltasseva, Alexandra and Cai, Wenshan and Liu, Yongmin},
  title = {Deep Learning for the Design of Photonic Structures},
  journal = {Nature Photonics},
  volume = {15},
  number = {2},
  pages = {77--90},
  year = {2021},
  doi = {10.1038/s41566-020-0685-y}
}

@article{Liu2018,
  author = {Liu, Dianjing and Tan, Yixuan and Khoram, Erfan and Yu, Zongfu},
  title = {Training Deep Neural Networks for the Inverse Design of Nanophotonic Structures},
  journal = {ACS Photonics},
  volume = {5},
  number = {4},
  pages = {1365--1369},
  year = {2018},
  doi = {10.1021/acsphotonics.7b01377}
}

@article{Chen2022WaveYNet,
  author = {Chen, Mingkun and Lupoiu, Robert and Mao, Chenkai and Huang, Der-Han and Jiang, Jiaqi and Lalanne, Philippe and Fan, Jonathan A.},
  title = {High Speed Simulation and Freeform Optimization of Nanophotonic Devices with Physics-Augmented Deep Learning},
  journal = {ACS Photonics},
  volume = {9},
  number = {9},
  pages = {3110--3123},
  year = {2022},
  doi = {10.1021/acsphotonics.2c00876}
}

@article{Lim2022MaxwellNet,
  author = {Lim, Joowon and Psaltis, Demetri},
  title = {{MaxwellNet}: Physics-Driven Deep Neural Network Training Based on {Maxwell}'s Equations},
  journal = {APL Photonics},
  volume = {7},
  number = {1},
  pages = {011301},
  year = {2022},
  doi = {10.1063/5.0071616}
}

@article{Augenstein2023,
  author = {Augenstein, Yannick and Rep{\"a}n, Taavi and Rockstuhl, Carsten},
  title = {Neural Operator-Based Surrogate Solver for Free-Form Electromagnetic Inverse Design},
  journal = {ACS Photonics},
  volume = {10},
  number = {5},
  pages = {1547--1557},
  year = {2023},
  doi = {10.1021/acsphotonics.3c00156}
}

@article{Pestourie2020ActiveLearning,
  author = {Pestourie, Rapha{\"e}l and Mroueh, Youssef and Nguyen, Thanh V. and Das, Payel and Johnson, Steven G.},
  title = {Active Learning of Deep Surrogates for {PDEs}: Application to Metasurface Design},
  journal = {npj Computational Materials},
  volume = {6},
  pages = {164},
  year = {2020},
  doi = {10.1038/s41524-020-00431-2}
}

@article{Pestourie2023PEDS,
  author = {Pestourie, Rapha{\"e}l and Mroueh, Youssef and Rackauckas, Chris and Das, Payel and Johnson, Steven G.},
  title = {Physics-Enhanced Deep Surrogates for Partial Differential Equations},
  journal = {Nature Machine Intelligence},
  volume = {5},
  number = {12},
  pages = {1458--1465},
  year = {2023},
  doi = {10.1038/s42256-023-00761-y}
}

@inproceedings{Li2021FNO,
  author = {Li, Zongyi and Kovachki, Nikola and Azizzadenesheli, Kamyar and Liu, Burigede and Bhattacharya, Kaushik and Stuart, Andrew and Anandkumar, Anima},
  title = {Fourier Neural Operator for Parametric Partial Differential Equations},
  booktitle = {International Conference on Learning Representations},
  year = {2021},
  url = {https://openreview.net/forum?id=c8P9NQVtmnO}
}

@article{Lu2021DeepONet,
  author = {Lu, Lu and Jin, Pengzhan and Pang, Guofei and Zhang, Zhongqiang and Karniadakis, George Em},
  title = {Learning Nonlinear Operators via {DeepONet} Based on the Universal Approximation Theorem of Operators},
  journal = {Nature Machine Intelligence},
  volume = {3},
  number = {3},
  pages = {218--229},
  year = {2021},
  doi = {10.1038/s42256-021-00302-5}
}

@article{Kovachki2023NeuralOperator,
  author = {Kovachki, Nikola and Li, Zongyi and Liu, Burigede and Azizzadenesheli, Kamyar and Bhattacharya, Kaushik and Stuart, Andrew and Anandkumar, Anima},
  title = {Neural Operator: Learning Maps Between Function Spaces with Applications to {PDEs}},
  journal = {Journal of Machine Learning Research},
  volume = {24},
  number = {89},
  pages = {1--97},
  year = {2023},
  url = {https://www.jmlr.org/papers/v24/21-1524.html}
}

@article{Raissi2019,
  author = {Raissi, M. and Perdikaris, P. and Karniadakis, G. E.},
  title = {Physics-Informed Neural Networks: A Deep Learning Framework for Solving Forward and Inverse Problems Involving Nonlinear Partial Differential Equations},
  journal = {Journal of Computational Physics},
  volume = {378},
  pages = {686--707},
  year = {2019},
  doi = {10.1016/j.jcp.2018.10.045}
}

@misc{Mao2024,
  author = {Mao, Chenkai and Lupoiu, Robert and Dai, Tianxiang and Chen, Mingkun and Fan, Jonathan A.},
  title = {Towards General Neural Surrogate Solvers with Specialized Neural Accelerators},
  year = {2024},
  eprint = {2405.02351},
  archivePrefix = {arXiv},
  primaryClass = {cs.LG},
  url = {https://arxiv.org/abs/2405.02351}
}

@article{Mao2026,
  author = {Mao, Chenkai and Fan, Jonathan A.},
  title = {Accurate and Scalable Deep {Maxwell} Solvers},
  journal = {Proceedings of the National Academy of Sciences},
  volume = {123},
  number = {18},
  pages = {e2530330123},
  year = {2026},
  doi = {10.1073/pnas.2530330123}
}

@inproceedings{Rizzuti2019Helmholtz,
  author = {Rizzuti, Gabrio and Siahkoohi, Ali and Herrmann, Felix J.},
  title = {Learned Iterative Solvers for the {Helmholtz} Equation},
  booktitle = {81st EAGE Conference and Exhibition 2019},
  pages = {1--5},
  year = {2019},
  doi = {10.3997/2214-4609.201901542}
}

@article{Azulay2022HelmholtzPreconditioner,
  author = {Azulay, Yael and Treister, Eran},
  title = {Multigrid-Augmented Deep Learning Preconditioners for the {Helmholtz} Equation},
  journal = {SIAM Journal on Scientific Computing},
  volume = {45},
  number = {3},
  pages = {S127--S151},
  year = {2023},
  doi = {10.1137/21M1433514}
}

@article{Lerer2024HelmholtzPreconditioner,
  author = {Lerer, Bar and Ben-Yair, Ido and Treister, Eran},
  title = {Multigrid-Augmented Deep Learning Preconditioners for the {Helmholtz} Equation Using Compact Implicit Layers},
  journal = {SIAM Journal on Scientific Computing},
  volume = {46},
  number = {5},
  pages = {S123--S144},
  year = {2024},
  doi = {10.1137/23M1583302}
}

@inproceedings{Goodfellow2014,
  author = {Goodfellow, Ian J. and Shlens, Jonathon and Szegedy, Christian},
  title = {Explaining and Harnessing Adversarial Examples},
  booktitle = {International Conference on Learning Representations},
  year = {2015},
  url = {https://arxiv.org/abs/1412.6572}
}

@inproceedings{Madry2018,
  author = {Madry, Aleksander and Makelov, Aleksandar and Schmidt, Ludwig and Tsipras, Dimitris and Vladu, Adrian},
  title = {Towards Deep Learning Models Resistant to Adversarial Attacks},
  booktitle = {International Conference on Learning Representations},
  year = {2018},
  url = {https://openreview.net/forum?id=rJzIBfZAb}
}

@techreport{Settles2009,
  author = {Settles, Burr},
  title = {Active Learning Literature Survey},
  institution = {University of Wisconsin--Madison, Department of Computer Sciences},
  number = {1648},
  year = {2009}
}

@inproceedings{Brinker2003,
  author = {Brinker, Klaus},
  title = {Incorporating Diversity in Active Learning with Support Vector Machines},
  booktitle = {Proceedings of the Twentieth International Conference on Machine Learning},
  pages = {59--66},
  publisher = {AAAI Press},
  year = {2003}
}

@inproceedings{Shrivastava2016OHEM,
  author = {Shrivastava, Abhinav and Gupta, Abhinav and Girshick, Ross},
  title = {Training Region-Based Object Detectors with Online Hard Example Mining},
  booktitle = {Proceedings of the IEEE Conference on Computer Vision and Pattern Recognition},
  pages = {761--769},
  year = {2016},
  doi = {10.1109/CVPR.2016.89}
}

@article{Lin1992ExperienceReplay,
  author = {Lin, Long-Ji},
  title = {Self-Improving Reactive Agents Based on Reinforcement Learning, Planning and Teaching},
  journal = {Machine Learning},
  volume = {8},
  number = {3--4},
  pages = {293--321},
  year = {1992},
  doi = {10.1007/BF00992699}
}

@inproceedings{Schaul2016PrioritizedReplay,
  author = {Schaul, Tom and Quan, John and Antonoglou, Ioannis and Silver, David},
  title = {Prioritized Experience Replay},
  booktitle = {International Conference on Learning Representations},
  year = {2016},
  url = {https://arxiv.org/abs/1511.05952}
}

@inproceedings{Ronneberger2015UNet,
  author = {Ronneberger, Olaf and Fischer, Philipp and Brox, Thomas},
  title = {{U-Net}: Convolutional Networks for Biomedical Image Segmentation},
  booktitle = {Medical Image Computing and Computer-Assisted Intervention -- MICCAI 2015},
  pages = {234--241},
  year = {2015},
  doi = {10.1007/978-3-319-24574-4_28}
}

@inproceedings{Long2015FCN,
  author = {Long, Jonathan and Shelhamer, Evan and Darrell, Trevor},
  title = {Fully Convolutional Networks for Semantic Segmentation},
  booktitle = {Proceedings of the IEEE Conference on Computer Vision and Pattern Recognition},
  pages = {3431--3440},
  year = {2015},
  doi = {10.1109/CVPR.2015.7298965}
}

@inproceedings{Kingma2015Adam,
  author = {Kingma, Diederik P. and Ba, Jimmy},
  title = {{Adam}: A Method for Stochastic Optimization},
  booktitle = {International Conference on Learning Representations},
  year = {2015},
  url = {https://arxiv.org/abs/1412.6980}
}

@article{Eckert2019,
  author = {Eckert, Regina and Chen, Michael and Wu, Fan and Waller, Laura and Ren, David and Chowdhury, Shwetadwip and Repina, Nicole},
  title = {High-Resolution {3D} Refractive Index Microscopy of Multiple-Scattering Samples from Intensity Images},
  journal = {Optica},
  volume = {6},
  number = {9},
  pages = {1211--1219},
  year = {2019},
  doi = {10.1364/OPTICA.6.001211}
}


\end{document}